\documentclass[10pt]{article}

\usepackage{epsfig}
\usepackage{bm}
\newcommand{\beq}{\begin{equation}}
\newcommand{\beqar}{\begin{eqnarray}}
\newcommand{\eeq}{\end{equation}}
\newcommand{\eeqa}{\end{eqnarray}}

\newcommand{\re}{{\sf I\!R}}
\newcommand{\me}{{\sf I\!M}}
\newcommand{\une}{{\sf 1\!I}}

\newcounter{llista}

\begin{document}

\title{Comment on ``Canonical formalism for Lagrangians with nonlocality of
finite extent"}

\author{Josep Llosa\\
Departament de F\'{\i}sica Fonamental, Universitat de Barcelona,\\
Diagonal 647, E-08028 Barcelona, Spain }

\date{\today}

\maketitle

\begin{abstract}
In ref. \cite{Woodard2000} it is claimed to have proved that Lagrangian theories with a nonlocality of finite extent are necessarily unstable. In this short note we show that this conclusion is false.
\end{abstract}


\section{Introduction}
In ref.\cite{Woodard2000} a canonical formalism for nonlocal Lagrangians
with a nonlocality of finite extent is established. It is compared with
Ostrogradski formalism \cite{Ostrogr} for local Lagrangians which depend
on a finite number of derivatives of coordinates. One of its central
conclusions is that Lagrangian systems with a nonlocality of finite
extent have no ``\ldots possible phenomenological role [\ldots] . They
have inherited the full Ostrogradskian instability \ldots".

The aim of the present note is:
\begin{list}
{(\roman{llista})}{\usecounter{llista}}
\item to point out some defects  in ref. \cite{Woodard2000} (section \ref{S2}) concerning the application of the variational principle that underlies the derivation of the nonlocal equation (2) from a Lagrangian; 
\item to stress the importance of the functional space where the variational
problem is developed, this is also the functional space where the
solutions must be searched (section \ref{S3}) and
\item to illustrate by two simple counterexamples (section \ref{S4}) that: {(A)}
Lagrangian systems containing derivatives of a higher order than first
are not necessarily unstable and {(B)} nonlocality of finite extent
does not inevitably lead to instability. 
\end{list}

\section{The nonlocal action principle \label{S2}}
Although the canonical formalism set up in ref. \cite{Woodard2000} is
derived on a general ground, it is basically illustrated by the simple
nonlocal Lagrangian system
\beq
L[q](t) = \frac12 m \dot{q}^2\left(t+\frac{\Delta}2\right) - \frac12 m
\omega^2 q(t) \,q(t+\Delta)
\label{e1}
\eeq
and the equation of motion for this Lagrangian is written as:
\begin{eqnarray}
\lefteqn{\int^{\Delta}_0 \frac{\delta L[q](t-r)}{\delta q(t)}\, dr =} \nonumber \\
 && - m\left\{ \ddot{q}(t)
+ \frac12 \omega^2 q(t+\Delta) + \frac12 \omega^2 q(t-\Delta) \right\} = 0
\label{e2}
\end{eqnarray}

It must be noticed that the latter equation as it reads does not properly
correspond to a standard action principle. Indeed, the action integral
whose variation would be the left hand side of (\ref{e2}) is:
\beq
S([q],t) = \int^{\Delta}_0 dr\,L[q](t-r) = \int^t_{t-\Delta} d\tau\, L[q](\tau) 
\label{e3}
\eeq
and equation (\ref{e2}) is equivalent to
\beq
\frac{\delta S([q],t)}{\delta q(t)} = 0
\label{e4}
\eeq
where $t$ is the same both in the numerator and the denominator. However,
the Euler-Lagrange equation that follows from the action principle $\delta
S = 0$ is 
$$ \frac{\delta S([q],t)}{\delta q(t^\prime)} = 0 \,, \qquad
\forall t^\prime $$ 
which is much more restrictive than (\ref{e4}).

Moreover, an equation like
\beq
- m\left\{ \ddot{q}(t)
+ \frac12 \omega^2 q(t+\Delta) + \frac12 \omega^2 q(t-\Delta) \right\} = 0 \,,
\label{e4a}
\eeq
 valid for $-\infty < t < \infty$,
cannot be derived from an action integral like (\ref{e3}), extending over a finite interval. Indeed, the variation of the action (\ref{e3}) is:
\begin{eqnarray}
\lefteqn{\delta S([q],t) = \left[ m \dot{q}\left(\tau+\frac{\Delta
t}2\right)\,\delta q\left(\tau+\frac{\Delta}2\right)\right]_{t-\Delta}^t -}  \nonumber \\
& & m \int_{t-\Delta}^t d\tau \left[\ddot{q}\left(\tau+\frac{\Delta
t}2\right)\, \delta q\left(\tau+\frac{\Delta}2\right) + \right.\nonumber \\
& & \left.\frac{\omega^2}2 q(\tau + \Delta) \delta q(\tau) + \frac{\omega^2}2 q(\tau) \delta
q(\tau+\Delta)\right] 
\label{e6}
\end{eqnarray}
The extremal condition $\delta S = 0$ then leads to the boundary conditions $\, \delta q(t+\frac{\Delta}{2}) =
\delta q(t-\frac{\Delta}{2}) = 0 \,$ and to the equations of motion:
\beq
\left. \begin{array}{lll}
(a) & \,t-\Delta <\tau < t-\frac{\Delta}{2}  & \quad  q(\tau+\Delta) = 0 \\
(b) & \,t-\frac{\Delta}{2}<\tau<t & \quad  
     \ddot{q}(\tau)+\omega^2/2 q(\tau+\Delta) = 0 \\
(c) & \,t <\tau<t+\frac{\Delta}{2} & \quad  
     \ddot{q}(\tau)+\omega^2/2 q(\tau-\Delta) = 0 \\
(d) & \,t+\frac{\Delta}{2}<\tau<t+\Delta &\quad  q(\tau-\Delta) = 0
\end{array} \right\}
\label{e7}
\end{equation}
which has the only solution $q(\tau) = 0\,$, for $t-\Delta< \tau<t
+\Delta$, as it ineluctably follows from sequentially exploiting (d), (b),
(a) and (c).

Furthermore, if we alternatively try with an action extended over a larger
interval:
$$ S=\int_0^T d\tau \,L[q](\tau)  $$
the Euler-Lagrange equations are:
\beq
\left. \begin{array}{ll}
(i)   &  q(\tau+\Delta) = 0 \\
(ii)  & 
     \displaystyle{\ddot{q}(\tau)+\frac{\omega^2}2 q(\tau+\Delta) = 0} \\
(iii)  & 
     \displaystyle{\ddot{q}(\tau)+\frac{\omega^2}2 [q(\tau+\Delta)+q(\tau-\Delta)] = 0 }\\
(iv) & 
     \displaystyle{\ddot{q}(\tau)+\frac{\omega^2}2 q(\tau-\Delta) = 0} \\    
(v)\quad & q(\tau-\Delta) = 0 
\end{array} \right\}
\label{e8}
\eeq
where the domains $(i)$ to $(v)$ respectively correspond to: $0 <\tau < \frac{\Delta}{2}$;  $\frac{\Delta}{2}<\tau<\Delta $; $\Delta <\tau< T$; $T <\tau<T+\frac{\Delta}{2}$ and $T+\frac{\Delta}{2}<\tau<T+\Delta$.
Equation (\ref{e8}) only looks like (\ref{e4a}) in the interval $\Delta<\tau<T$. 

The conditions (\ref{e8}.i) and (\ref{e8}.v) then yield:
$$ q(\tau) = 0 \, ; \qquad \Delta<\tau < 3\,\frac{\Delta}2 \quad {\rm or} \quad T - \frac{\Delta}2 <\tau <T $$
that act as constraints on the possible solutions of (\ref{e8}.ii),
(\ref{e8}.iii) and (\ref{e8}.iv). As a consequence, equations (\ref{e8}) 
can be reduced to an ordinary differential equation, whose order depends
on the number of times that the elementary length $\Delta$ fits into
$[0,T]$.

We have thus illustrated the important role played by the integration 
bounds in the nonlocal action (\ref{e3}) as far as the Euler-Lagrange
equations are concerned. The integration bounds in the action and the
problems associated to them are commonly overlooked in theoretical physics
literature because, in standard local cases no trouble is usually entailed
by proceeding in this manner. Nonlocal cases are however a new ground where nothing
can be taken for granted.

For a local action, the bounds of the integral also determine the
functional Banach space where the variational calculus is meaningful
\cite{Gelfand}, e. g., the space ${\cal C}^2([a,b])$ for an action
integral extending over $[a,b]$. This is also the space where the
solutions to the Euler-Lagrange equations have to be sought.

A way to derive equation (\ref{e2}), for $t$ extending from $-\infty$ to
$\infty$, from an action principle could consist in taking the integral
over the whole $\re$:
\beq
 S = \int_{-\infty}^\infty d\tau\, L[q](\tau)  \,,
\label{e10}
\eeq
but then two additional difficulties arise: on the one hand, the action
$S$ does not converge anymore for all $q\in {\cal C}^2(\re)$ and, on the
other, ${\cal C}^2(\re)$ is not a Banach space. (The variational calculus
should be then approached in terms of Fr\'echet spaces \cite{Hamilton},\cite{dones}.)
To my knowledge, it remains an open problem to establish the appropriate mathematical framework where a nonlocal equation like (\ref{e4a}) can be derived from an action integral like (\ref{e10}). This results in a lack of preciseness in the definition of the
functional space where the nonlocal equation has to be solved.

\section{The stability problem \label{S3}}
Leaving aside the difficulties just mentioned, suppose that, for some physical reasons whatsoever, we are only interested on the solutions of (\ref{e4a}) in the Banach space
$$ {\cal B} = \{ q\in {\cal C}^2(\re) \,; \;|q(t)|\,,\;|\dot{q}(t)|\;\;  {\rm and}\;\; |\ddot{q}(t)|\; \mbox{ are bounded}\} \,.$$
The general solution of (\ref{e8}) is thus
\beq
q(t) = \sum_l \left(A_l e^{ik_lt}+A^\ast_{l} e^{-ik_lt}\right)
\label{e9}
\eeq
where $\pm k_l$ are the real solutions 
\footnote{A complex value of $k_l$ would result in an exponential growth either at $+\infty$ or $-\infty$ and then $q\notin{\cal B}$}
 of
\beq
h(k) \equiv k^2 - \omega^2\cos(k \Delta) = 0
\label{e11}
\eeq
and $A^\ast_l$ is the complex conjugate of $A_l$, to ensure that $q(t)\in \re$.

\begin{figure}[htbp]
        \begin{center}
                \epsfig{file=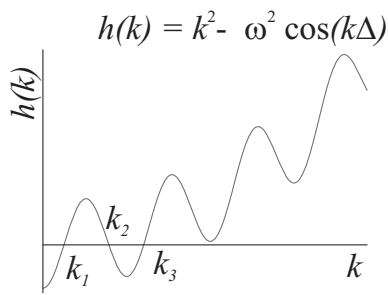,width=5cm}
        \end{center}
        \caption{$h(k)=0$ has a finite number of real roots, and the sign of the derivative $h^\prime(k_l)$ at each root is alternating. }
        \label{f1}
\end{figure}

Notice that the number of real roots of (\ref{e11}) is finite. A look
at figure \ref{f1} is enough to get convinced that they can be indexed so
that
$$k_j<k_i \quad{\rm if} \quad j<i  \quad {\rm and}\quad 
l=1,2,\ldots N \,.$$ 
$N$ can be either odd and then all roots are simple, or even, in which case
$\pm k_N$ are both double. It should also be remarked that the greater is $\Delta$, the denser is the wiggling in the graphics (figure \ref{f1}). Therefore, $N$ increases with $\Delta$.

The space of solutions of (\ref{e4a}) in ${\cal B}$ can be hence
coordinated by $\,2N<\infty\,$ real parameters, namely, the real and imaginary 
parts of $A_l$, that can be put in correspondence with
the initial data: $q_0$, $\dot{q}_0$, \ldots $q^{(2N-1)}_0$.

The solutions of (\ref{e4a}) in ${\cal B}$ are stable because a small
change in the initial data $\delta q_0^{(\alpha)}\,$,  
results in a small change in the complex parameters:
$\delta A_l\,$. Indeed, from the linearity of
eq. (\ref{e4a}) and from the general solution (\ref{e9}) it follows that:
$$ \sum_{l=1}^N\left[\delta A_l\,(ik_l)^\alpha + \delta A_l^\ast
(-ik_l)^\alpha \right] = \delta q_0^{(\alpha)} \,, $$
$\alpha=0,1,\ldots, 2N-1 \,$,
which can be inverted to obtain $\delta A_l$ as a linear function of
$\delta q_0^{(\beta)}$. Therefore, there exists $K>0$ such that
$\left|\delta A_l\right| \leq K ||\delta q_0||\,$, where $||\delta
q_0||\equiv \sup\{|\delta q_0^{(\alpha)}|\,; \; \alpha =0,1,\ldots
2N-1\}\,$.  The deviation from $q(t)$ evolves with time as: 
$$ |\delta q(t)|=\left|\sum_{l=1}^N \Re\left(\delta A_l e^{ik_lt}\right)\right|
\leq \sum_{l=1}^N 2\left|\delta A_l\right| \leq 2NK ||\delta q_0|| $$ 
which proves the stability
\footnote{In the sense of Liapounov, see \cite{Pontria}} 
of the solutions of (\ref{e4a}) in the space ${\cal B}$.

Notwithstanding, if we now have a look at the Hamiltonian (equation (48)
in ref. \cite{Woodard2000}):
\begin{eqnarray*}
H(t) &=& \frac12 m \dot{q}^2(t) + \frac 12 m \omega^2 q(t) q(t+\Delta) - \\
     & & \frac 12 m \omega^2 \int_0^{\Delta} ds \dot{q}(t+s) q(t+s-\Delta) \,,
\end{eqnarray*}
on substituting the general solution (\ref{e9}), we obtain that
\beq
H(t) = 2 m\,\sum_{l=1}^N k_l g(k_l)\,A_l A_l^\ast 
\label{e12}
\eeq
with 
\beq
g(k) = k + \frac{\omega^2}2 \,\Delta\, \sin(k\Delta)
\label{e13}
\eeq
As expected, $H(t)$ is an integral of motion, but it has not a definite
sign. Indeed, notice that $g(k)=h^\prime(k)/2$ alternates sign at each root $k_l$ [see eq. (\ref{e11}) and figure \ref{f1}]. Therefore, $g(k_l)$ is positive or
negative depending on whether $l$ is even or odd, respectively (moreover,
$g(k_l)=0$ if $k_l$ is a double root).


\section{Two simple counterexamples \label{S4}}
\subsection{The so-called Ostrogradskian instability}
In ref. \cite{Woodard2000} it is proved that the Hamiltonian formalism for
a nonlocal Lagrangian can be obtained as a limit case for
$N\rightarrow\infty$ of the Ostrogradski formalism \cite{Ostrogr} for a
Lagrangian that depends on the derivatives of the coordinates up to order
$N$.  
\footnote{A similar result was also obtained in Ja\'en, X., J\'auregui, R., Llosa, J. and Molina, A., {Phys. Rev. D} {\bf 36}, 2385 (1987)}
For $N>1$, the Ostrogradski Hamiltonian is linear on all the
canonical momenta but one, namely, $P_1$, \ldots $P_{N-1}$, therefore it
has not a definite sign. The fact that the energy is not bounded from
below is then argued to conclude that the solutions of the equations of
motion are ineludibly unstable. This is what is called the {\it
Ostrogradskian instability}. It is also shown in \cite{Woodard2000} that
this drawback also holds in the limit $N\rightarrow\infty$.

Actually what has been proved there is only that the energy cannot be
taken as a Liapunov function \cite{Boyce} to conclude the stability of the
equations of motion derived from an $N^{\rm th}$ order Lagrangian ($N>1$).
However, the fact that a sufficient condition of stability is not met does not
imply instability. Let us consider the following simple counterexample: 
\beq
L(q,\dot{q},\ddot{q}) = \frac12 \ddot{q}^2 + \frac12 B \dot{q}^2 + \frac12
C {q}^2  
\label{h2}
\eeq
where $B$ and $C$ are two parameters which we shall later tune in order to
get stability.

According to Ostrogradski theory, the canonical coordinates and momenta
are [in the notation of ref. \cite{Woodard2000}, eqs. (6-7)]
$$ Q_1 = q \,, \qquad Q_2 = \dot{q} \,\qquad
 P_1 = B \dot{q} - q^{(iii)} \,, \qquad P_2 = \ddot{q} \, $$
and the Hamiltonian is (eq.(9) in \cite{Woodard2000}):
\beq 
H = \frac12 P_2^2 + P_1 Q_2 - \frac12 B Q^2_2 - \frac12 C Q_1^2
\label{h3}
\eeq

Introducing 
$$ \vec X =\left(\begin{array}{c} Q_1 \\ Q_2 \\ P_1\\ P_2 
                                        \end{array} \right)
                                        \qquad {\rm and } \qquad
         \me = \left(\begin{array}{cccc} 
                                0 & 1 & 0 & 0 \\
                                0 & 0 & 0 & 1 \\ 
                                C & 0 & 0 & 0\\
                                0 & B & -1 & 0
                                \end{array} \right)     $$
the Hamilton equations for (\ref{h3}) can be then written as the linear system:
$$ \frac{d\;\;}{dt} \, \vec X = \me \vec X \,,$$
the stability of whose solutions depends on the real part of the roots of
the characteristic polynomial $p_\me(\lambda) = \det(\me-\lambda \une_4)$, that
is:
$$ \lambda =\pm \sqrt{\frac{B\pm\sqrt{B^2-4 C}}2} $$
If the parameters are tuned so that $B<0$ and $0<C<B^2/4$, then all 
roots are imaginary and the system is stable \cite{Boyce}. The latter is
not an obstacle for the fact that the Hamiltonian has not a definite sign.

\subsection{A case of finite extent nonlocality}
In the next example, the boundaries of the action integral are finite. 
Consider the nonlocal action $S[q]=\int_0^T dt\,L[q](t)$, with
\beq
L[q](t) = \frac12 \dot{q}^2(t) - \frac12 \omega^2 q^2(t) +
\frac{\omega^4}2 q(t)\, \int_0^T dt^\prime \,G(t,t^\prime)\,q(t^\prime)
\label{l1}
\eeq
where, for $(t,t^\prime)\in[0,T]^2$,
\begin{eqnarray}
G(t,t^\prime)&=& \frac{-1}{\omega \sin \omega T} 
\left[\sin\omega(T-t^\prime) \,\sin\omega t \,\theta(t^\prime-t) +
\right.\nonumber \\ 
  && \hspace*{4.5em}\left.\sin\omega(T-t) \,\sin\omega t^\prime
\,\theta(t-t^\prime)\right] 
\label{l2}
\end{eqnarray}
and is the solution of:
\beq
\partial_t^2 G(t,t^\prime) + \omega^2 G(t,t^\prime) =\delta(t-t^\prime) 
\label{l3}
\eeq
for the boundary conditions: $G(0,t^\prime)=G(t,T)=0$.

The variation $\delta S =0$ with the boundary conditions $\delta q(0) =
\delta q(T) = 0$ leads to the equations of motion
\beq
\ddot{q}(t) + \omega^2 q(t) - \omega^4 \int_0^T dt^\prime\,
G(t,t^\prime)\, q(t^\prime) = 0
\label{l4}
\eeq
The solutions $q(t)$ must be sought in the Banach space ${\cal C}^2([0,T])$.

Differentiating twice (\ref{l4}) and taking (\ref{l3}) into account, we
arrive at:
\beq
q^{(iv)} + 2 \omega^2\,\ddot{q} = 0
\label{l5}
\eeq
Hence, the solutions of (\ref{l4}) must be among the general solution of
(\ref{l5}):
\beq
q(t) = A e^{i\alpha t} + A^\ast e^{-i\alpha t} + D t + E
\label{l6}
\eeq
with $\alpha = \omega\sqrt{2}$. The parameters $A$, $A^\ast$, $D$ and $E$
must fulfill the following constraints 
$$ D=0 \qquad {\rm and} \qquad E = A + A^\ast $$
which result from substituting (\ref{l6}) into (\ref{l4}).

The general solution of (\ref{l4}) is therefore:
\beq
q(t) = A \left(e^{i\alpha t} + 1\right) + A^\ast \left(e^{-i\alpha t} +
1\right)
\label{l7}
\eeq
The phase space for our system is thus two-dimensional, and every solution
is determined by the initial values $q_0$ and $\dot{q}_0$:
$$ q_0 = 2 (A+A^\ast) \qquad {\rm and} \qquad \dot{q}_0 = i\alpha (A-A^\ast)$$

By direct inspection of (\ref{l7}), we see that the solutions of equation 
(\ref{l4}) are stable, although the latter is derived from a Lagrangian with a nonlocality of finite extent. That is, for any $\epsilon>0$ there exists $\rho>0$
such that $|\delta q_0|+ |\delta \dot{q}_0|< \rho$ implies that $|\delta
q(t)|+ |\delta \dot{q}(t)|< \epsilon$, for all $t$, which proves the stability.

\section{Conclusion}
We have intended to stress the crucial importance of clearly precising the Banach space where the variational principle for a nonlocal Lagrangian is formulated. This degree of precision is usually obviated in theoretical physics (i. e., for local Lagrangians) without any major problem. However, such non rigourous way of proceeding cannot be extrapolated to systems with a new complexity. The relevance of the above mentioned Banach space is twofold: (i) it is where the solutions of the equations of motion must be sought and (ii) it is the function space where path-integrals are to be calculated in an eventual quantization of the system.

We have also analised the stability of the equations of motion for a Lagrangian system presenting a nonlocality of finite extent. We have shown that the choice of the Banach space where the variational principle is meaningfully formulated is crucial to decide the stability or unstability of the system. Furthermore, we have seen that a system can be stable in spite of the fact that the Hamiltonian does not have a minimum. 

Finally, we have shown by a counterexample that higher order Lagrangian systems are not necessarily unstable. The fact that a sufficient condition for stability is not fulfilled does not imply instability.

\end{document}